\documentclass[12pt]{article}
\renewcommand{\theequation}{\arabic{section}.\arabic{equation}}
\newcounter{saveeqn}

\newcommand{\reseteqn}{\setcounter{equation}{\value{saveeqn}}%
\renewcommand{\theequation}{\arabic{section}.\arabic{equation}}}
 \reseteqn

\begin{document}

\begin{titlepage}

\begin{center}{\Large \bf Optimal covariant fitting  to a Robertson-Walker 
metric and smallness of backreaction} \\[5 ex]

{\bf Dieter Gromes}\\[3 ex]Institut f\"ur
Theoretische Physik der Universit\"at Heidelberg\\ Philosophenweg 16,
D-69120 Heidelberg \\ E - mail: d.gromes@thphys.uni-heidelberg.de \\
 \end{center} \vspace{2cm}

{\bf Abstract:} We define a class of ``optimal''  coordinate systems
by requiring that the deviation from an exact Robertson-Walker metric 
is ``as small as possible'' within a given four dimensional volume. The
optimization is performed by minimizing several volume integrals
which would vanish for an exact Robertson-Walker metric. 
Covariance is automatic.
Foliation of space-time is part of the optimization procedure. 
Only the metric is involved in the procedure, no assumptions about 
the origin of the energy-momentum tensor are needed. 
A scale factor does not show up during the optimization process,
the optimal scale factor is determined at the end. The general
formulation is non perturbative. An explicit perturbative treatment is 
possible. The shifts which lead to the optimal coordinates obey 
 Euler-Lagrange equations which are formulated and solved in first order of the 
perturbation.
The extension to second order is sketched, but turns out to be unnecessary.
The only freedom in the choice of coordinates which finally remains  are the
rigid transformations
which keep the form of the Robertson-Walker metric intact, i.e. 
translations in space and time, spatial rotations, and spatial scaling. 
Spatial averaging becomes trivial. 
In first order of the perturbation there is no backreaction. 
A simplified second order treatment results in a very small effect, excluding 
the possibility to mimic
 dark energy from backreaction. This confirms (as well as 
contradicts) statements in the literature.

\vfill \centerline{November 2011}

\end{titlepage}

\section{Introduction}

The averaging problem, i.e. the problem of averaging a realistic 
inhomogeneous metric 
into a smooth one, as well as the fitting problem, the fitting 
of an ``optimal'' 
Robertson-Walker (R-W) metric to a realistic inhomogeneous metric, are both 
non trivial due to the freedom of choosing arbitrary coordinates 
in general relativity. Most papers focus on ``gauge transformations'', 
where the R-W background metric $\overline{g}_{\mu\nu}$ is given and fixed, 
and only the perturbation $h_{\mu\nu}$ is transformed. To fix the 
background one usually resorts to a flow of matter. 
This is unsatisfactory for two reasons. Firstly one has to make
rather stringent assumptions concerning the flow, like only one single 
component and
absence of rotation. The second point, although quite 
obvious but nevertheless hard to find  mentioned in the literature, 
appears even more drastic: 
The flow which is used is, of course, not the real flow of matter but
already some average over an irregular flow. The result
of this averaging clearly depends on the choice of coordinates. 
One is faced with the bizarre situation that one starts some sophisticated 
``gauge invariant'' averaging procedure on the basis of a background obtained 
from an ambiguous and unspecified averaging. This makes the 
whole procedure quite dubious. 
 
There is an extensive literature on both topics 
which can only be briefly addressed  here. We refer e.g. to the monograph of 
Krasi\'nski \cite{Kras} and the comprehensive review of 
Buchert \cite{Buchert}. A careful analysis of the fitting problem was 
given by Ellis and Stoeger \cite{EllisStoeger}. 
We only briefly mention some aspects here.
The averaging problem was first raised by 
Shirokov and Fisher \cite{ShirFi} in 1963.   The authors
suggested to
integrate the metric tensor over a four dimensional volume with the 
familiar factor $\sqrt{-g}$ in the measure. Such an expression is, 
however, not covariant for a tensor due to the freedom of 
performing local transformations.
A covariant averaging prescription can be constructed by introducing a  
bivector $g_\alpha^{\beta }(x,x')$ of 
geodesic parallel displacement, as discussed 
in the appendix of \cite{Isaac}. This
transforms as a vector with respect to coordinate transformations at
either $x$ or $x'$ and maps a vector $A_\beta (x')$ to 
$\bar{A}_\alpha(x)= g_\alpha^{\beta }(x,x') A_\beta(x')$, analogously for 
higher order tensors.
An  averaging with the help of bivectors is also
used in the work of Zalaletdinov \cite{Zalat1} where the emphasis was on the 
commutativity of averaging and covariant differentiation. As remarked by 
Stoeger, Helmi, and 
Torres \cite{StoeHT} the method of using a covariantly conserved 
bivector is not applicable to the metric, because 
the covariant derivative of the metric vanishes. The metric is therefore 
invariant under this averaging procedure. Another popular method due to 
Bardeen \cite{Bardeen} is to work 
with gauge invariant (in first order of the perturbation and for static 
transformations only) quantities.
The most general covariant and translation invariant first order 
averaging scheme has been given in \cite{Gromes}. But any such an 
averaging has the principal problem that a plane wave, instead of 
being averaged to zero, will always stay a plane wave,
albeit with reduced amplitude: 
$\exp(i{\bf k x}) \rightarrow 
\int f({\bf x} - {\bf y}) \exp(i{\bf k y}) d^3 y 
= \{\int f({\bf z}) \exp(-i{\bf k z}) d^3z \} \cdot \exp(i{\bf k x})$.
\\ 

Instead of attempting an averaging, it appears therefore more promising 
to determine directly an ``optimal'' approximating smooth metric.
Our approach is conceptually simple. We fix the coordinate system as far as 
principally possible, so that no unphysical gauge freedom remains. The
coordinate system is chosen in such a way that, in a given four
dimensional volume, the metric is as close to an exact R-W 
metric as possible. Before going into details one should recall that
an exact R-W metric (with $k=0$) keeps its
form under an eight parameter group of global symmetry transformations:
Rigid translations in
space and time, rigid spatial rotations, and rigid scaling of the
spatial coordinates. (For $k=\pm 1$ there
is only a seven parameter group, scaling is not allowed.)
In the case of translations in time, and of
scaling in space, the scale function $a(t)$ changes. 

This freedom in the choice of coordinates cannot and need not be fixed.
It is inevitably connected with the symmetry of the R-W metric. 
Any covariant fitting procedure will necessarily share this freedom 
of transformations. Therefore a maximal fixing of the coordinate
system means that the coordinates are fixed up to the above rigid 
transformations, while no further transformations are allowed anymore.
The coordinate system obtained at the end should
fulfill two criteria:
\\

{\bf Covariance:} 
Let two observers $A,\, B$ describe the same realistic inhomogeneous space 
in different and completely arbitrary coordinate systems 
$S_A$ and $S_B$. Both of them apply the same 
definite method to transform to ``optimized'' systems $S'_A$ and
$S'_B$ respectively. Then the  systems $S'_A$ and $S'_B$ thus obtained 
can only be
related by a transformation from the eight parameter group above. 
\\

{\bf Optimization:} 
The metric in the optimized system should be ``as close to an 
exact R-W metric as possible'' within a given four dimensional volume. 
Since any perturbative
treatment is performed around a R-W background, this requirement 
guarantees that the perturbation becomes as ``small'' as possible.
One has to define  the conditions of this optimization and to construct
the ``optimal'' coordinate system. 
\\

Our covariant optimization proceeds via
 a series of minimizations of four dimensional volume 
integrals. Expressions which
vanish for the exact R-W metric are minimized by choosing an
optimal gauge. This gauge fixing is performed as far as
principally possible. Starting from an arbitrary system $S$ with metric
$g_{\mu\nu}(x)$, one constructs an ''optimal'' system $S'$ with the  
transformed metric $g'_{\mu\nu}(x')$. At the end one can define  the ''optimal''
approximating R-W metric $\overline{g}_{\mu\nu}(x')$.

The method has the following properties and advantages.

\begin{itemize}
\item It only uses the metric $g_{\mu\nu}$, no assumptions about the 
origin of the energy momentum tensor  are necessary. A scale factor
does not show up during the procedure, the optimal scale factor 
is determined at the end.
\item Foliation of space-time is part of the procedure and is obtained 
in a unique way, again without resorting to any assumptions concerning 
a flow of matter.
\item The general formulation is non perturbative. 
\item The procedure can be explicitly applied in perturbation
theory if the metric is a small deviation from an exact R-W metric.
We will present the explicit formulae in first order of the 
perturbation and sketch the procedure for the second order. 
\item The four dimensional volume over which the minimization is performed 
is arbitrary. For the perturbative treatment we will specialize 
to simple volumes.
\item Covariance is an immediate consequence of the method. Since the 
variation is taken over all coordinate systems, it does not matter 
in which coordinate system one starts.
\item Spatial averages of arbitrary tensor fields can be naively 
performed when using the optimal coordinate system. There is no need 
to decompose into tensor structures or to restrict to 
averaging of scalar quantities.
\item There is no backreaction in first order.
\end{itemize}

The paper is organized as follows.
In sect. 2 we present the general non perturbative method. 
In sect. 3 
this is explicitly applied in first order of the perturbation. 
The Euler-Lagrange equations for the coordinate shift which leads to the 
optimal system are formulated and solved. Boundary effects turn out to 
be irrelevant if the wave length of the perturbation is small compared 
to the spatial extension of the volume. In sect. 4 we present the 
rather simple extension to second order.
The short sect. 5 describes averaging which has become trivial. 
Sect. 6 deals with backreaction. There is no
backreaction in first order. In a simplified static treatment of the 
second order it turns out that the second order of the transformation is 
not needed, only the first order perturbation of the metric introduced 
into the second order 
Einstein tensor enters. We consider the contributions of galaxy clusters, 
galaxies, and stars. We give arguments why neither a concentration of 
clusters and 
galaxies in bubble walls which surround large voids nor retardation effects
are relevant.
We find that the ratio $\rho_b/\overline{\rho}$ between the density $\rho_b$
mimicked by backreaction, and the averaged 
matter density  $\overline{\rho}$ is small, of the order 
of $10^{-4}$ to at most $10^{-2}$. 
To mimic dark energy from backreaction appears practically impossible.
Sect. 7 gives a summary.

\setcounter{section}{1}
\setcounter{equation}{0}\addtocounter{saveeqn}{1}%

\section{General conditions for the optimal coordinates}

For an exact spatially flat ($k=0$) Robertson-Walker (R-W) metric 
$\overline{g}_{\mu\nu}$ one has

\begin{equation}
 \overline{g}_{mn} = a^2 (t) \delta_{mn},  \; \overline{g}_{m0} = 0, 
\; \overline{g}_{00} = -1.
\end{equation}
Consider now a realistic metric with metric  
tensor $g_{\mu\nu}(x)$, and a given four dimensional volume. We want to 
define ``optimal'' coordinates $x'^\mu$ in which the metric within this 
volume becomes, in a
sense to be defined, as close as possible to an exact R-W
metric $\overline{g}_{\mu\nu}$ with $k=0$.
It would be impractical to combine all conditions into a single
variation problem by minimizing an integral over the sum of
appropriate squares. This would lead to rather complicated
Euler-Lagrange equations even in a perturbative
treatment. It is technically much simpler to proceed in steps.
Each of the first four steps approximizes a certain property of the exact 
R-W metric. A
scale factor does not show up in the conditions. The optimal scale 
factor $a(t)$ associated with the given realistic metric is determined at 
the end of the procedure in step 5. 

All integrals in the four steps below are taken over the four dimensional 
volume under consideration. In step 1 the time coordinate is fixed but 
arbitrary, and the variation is taken over all primed 
systems which  are time independent coordinate transformations of the 
original one. 
In steps 2 - 4 we also allow time dependent transformations. These 
transformations
have to respect the restrictions obtained in the previous steps.  
Because all coordinate systems are admitted in the variation procedure
the covariance of the method is automatically guaranteed. Each step 
restricts the freedom of choice of coordinates more and more, at the end the 
coordinates are fixed as far as principally possible.
Step 1 leads to a transversal perturbation. 
Usually one will start already with some ``reasonable'' coordinate system.
In this case steps 2, 3, 4 become trivial. 
In step 5, finally, we define the 
optimal scale factor  $a(t)$ which gives the ``best'' approximation of the 
given metric to an exact R-W metric. Here and in the following 

\begin{equation}
\langle f \rangle  
= \langle f \rangle (t) = 
\frac{\int f({\bf y},t) \sqrt{^3g({\bf y},t)}d^3y}
{\int \sqrt{^3g({\bf y},t)}d^3y}
\end{equation}
denotes the spatial average of $f({\bf x},t)$.

\begin{eqnarray}
{\bf Step \; 1: } & & 
\int \Bigg( \frac{g'_{mn}(x') -
\langle g'_{ii}/3 \rangle \delta_{mn}}{\langle g'_{jj}/3 \rangle} \Bigg)^2
\sqrt{-g'(x')}d^4x'  =  \mbox{ Minimum}, 
\nonumber\\
& & \mbox{where time is fixed but arbitrary, and the variation 
is over all time}
\nonumber\\
& &  \mbox{independent coordinate transformations.}
\\
{\bf Step  \; 2: } & &
\int \Bigg( \frac{g'_{m0}(x')}{\sqrt{\langle -g'_{00} \rangle
\langle g'_{jj}/3 \rangle}}  
\Bigg)^2 \sqrt{-g'(x')}d^4x' =  \mbox{ Minimum},
\nonumber\\
& & \mbox{with the variation taken over all coordinate transformations} 
\nonumber\\
& & \mbox{which respect the restrictions obtained in step 1.}
\\
{\bf Step  \; 3: } & & 
\int \Big( g'_{00}(x') + 1 \Big)^2
\sqrt{-g'(x')} d^4x' =  \mbox{ Minimum},
\nonumber\\ 
& & \mbox{with the variation taken over all coordinate transformations} 
\nonumber\\
& & \mbox{which respect the restrictions obtained in steps 1,2.}
\\
{\bf Step  \; 4: } & &
\int \Bigg( 
\frac{\partial}{\partial t'} g'_{00}(x')
\Bigg)^2 \sqrt{-g'(x')}d^4x'
 =  \mbox{ Minimum},
\nonumber\\
& & \mbox{with the variation taken over all coordinate transformations} 
\nonumber\\
& & \mbox{which respect the restrictions obtained in steps 1,2,3.}
\\
{\bf Step  \; 5: } & &
\mbox{Define the optimal scale factor $a(t')$ by }
a^2(t') \equiv \langle g'_{ii} /3\rangle (t'). 
\end{eqnarray}

Summation convention is always understood, also for identical 
lower indices.

The meaning of the conditions should be obvious. For an exact R-W metric 
all the integrands would vanish. In steps 1 and 2
we introduced normalization factors in the denominator.
This is necessary, because otherwise one would run into an unphysical
minimum by a simple scaling of the metric. 

The conditions above do not fix the metric completely.
They  yield
a whole class of optimal coordinate systems. This class is, by construction,
independent of the system with which one has started.
The freedom which remains are the transformations from the eight 
parameter group of rigid
translations in space and time, rigid rotations in space, and
rigid scaling of the space coordinates. This is just the
invariance group of coordinate transformations mentioned in the introduction 
which keep the form of the exact R-W metric intact. 

\setcounter{section}{2}
\setcounter{equation}{0}\addtocounter{saveeqn}{1}%

\section{First order}
We consider a perturbed Robertson-Walker metric of the form

\begin{equation}
g_{mn}(x) = \widetilde{a}^2 (t) \delta_{mn} + h_{mn}(x), \; 
g_{m0}(x) = h_{m0}(x), \; g_{00}(x) = -1+h_{00}(x).
\end{equation}
The gauge and the way of splitting into background and perturbation 
is completely arbitrary, except that the perturbation $h_{\mu\nu}$ 
should be small. The scale factor $\widetilde{a}(t)$ is in general 
not identical 
with the optimal scale factor $a(t)$ obtained at the end.

Introduce the 
perturbed metric in the primed system into the integrands 
in (2.3) - (2.6). The primed system is connected to the old one by 
an infinitesimal transformation  $x^\mu = {x'}^\mu + \xi ^\mu$.
Because all the brackets vanish for the unperturbed 
metric it is sufficient to expand these up to first  order in ${h'}_{\mu\nu}$
and $\xi_\mu$. Furthermore we can put 
$\sqrt{-g} = \widetilde{a}^3(t)$ and restrict to the leading order 
in the denominators.
Express the ${h'}_{\mu\nu}$ 
in the primed system by the $h_{\mu\nu}$ in the old one and the 
shifts $\xi_\mu$. In lowest order one has 
$\xi _m = \widetilde{a}^2(t) \xi^m, \;\xi_0 = - \xi^0$. The well 
known transformation 
laws for the metric in lowest order of $\xi_\mu$ are:

\begin{eqnarray}
g'_{mn}(x') & = & g_{mn}(x) +  \xi _m,_n + \xi _n,_m \nonumber\\
& = & \widetilde{a}^2(t) \delta_{mn} + h_{mn} +  \xi _m,_n + \xi _n,_m \\ 
& = & \widetilde{a}^2(t') \delta_{mn} + h_{mn} +  \xi _m,_n + \xi _n,_m 
- 2 \widetilde{a}(t') \dot{\widetilde{a}}(t')\xi_0 \delta_{mn}, 
\\
g'_{m0}(x') & = & g_{m0}(x) +  \xi _0,_m + \, \xi_{m,0} 
- 2 (\dot{\widetilde{a}}(t)/\widetilde{a}(t))\xi_m \nonumber\\ 
 & = & h_{m0} +  \xi _{0,m} + \widetilde{a}^2(t') \xi^m,_0,
\\
g'_{00}(x') & = & g_{00}(x) + 2 \xi  _0,_0 \nonumber\\
& = & -1+h_{00} + 2 \xi  _0,_0.
\end{eqnarray}
The transformation of the averages $\langle g'_{\mu\nu} \rangle (t')$, 
which now 
refer to a different time $t'$, is most easily obtained by writing 
$d^3y' = \delta ({y'}^0-t') d^4y'$. This results in 

\begin{eqnarray}
\langle g'_{ii} /3\rangle  & = & \widetilde{a}^2(t) 
+ \langle \frac{h_{ii}}{3} + \frac{2}{3} \xi_i,_i \rangle 
+ 2 \widetilde{a} \dot{\widetilde{a}} (\xi_0 - \langle \xi_0 \rangle )
\\
& = & \widetilde{a}^2(t') + \langle \frac{h_{ii}}{3} 
+ \frac{2}{3} \xi_i,_i \rangle 
- 2 \widetilde{a} \dot{\widetilde{a}}  \langle \xi_0 \rangle .
\end{eqnarray}
In this way we obtain the integrals which have to
be minimized. 
\\

{\bf Step 1:}

We have to minimize

\begin{eqnarray} & & \lefteqn{\int \Bigg( \frac{{g'}_{mn}(x') -
\langle {g'}_{ii}/3 
\rangle \delta_{mn}}{\langle {g'}_{jj} /3\rangle } \Bigg)^2
\sqrt{-g'(x')}d^4x' = }\\
& & \int \frac{1}{\widetilde{a}(t)} \Bigg( h_{mn} 
- \frac{1}{3} \langle h_{ii} \rangle
\delta_{mn}
 + \xi _m,_n + \xi _n,_m  - \frac{2}{3} \langle \xi_i,_i \rangle 
\delta_{mn} 
- 2 \widetilde{a}(t)\dot{\widetilde{a}}(t) (\xi_0 
- \langle \xi_0 \rangle)\delta_{mn}\Bigg) ^2 d^4x
\nonumber
\end{eqnarray}
with respect to $\xi_m$ while keeping $\xi_0$ arbitrary but fixed.
Here and in the following it is irrelevant whether we consider the 
expressions in the old or in the new 
system. In all brackets the leading terms cancel, changes in the 
boundary of the volume only contribute to higher order.

From a variation  $\delta \xi_m$ in the interior we  
obtain the Euler-Lagrange (E-L) equations

\begin{equation}
\xi_m,_{nn} + \xi_n,_{mn} +h_{mn},_n 
- 2 \widetilde{a}(t)\dot{\widetilde{a}}(t) \xi_{0,m} =0.
\end{equation}
We use the usual decompositions for $h_{mn}$ and $h_{m0}$:

\begin{eqnarray}
h_{mn}  & =  & \widetilde{a}^2(t) [A \delta_{mn} + B_{,mn} 
+ C_{m,n} + C_{n,m} + D_{mn}],
\\
h_{m0} & = & \widetilde{a}(t)[F,_m + G_m],
\end{eqnarray}
with
\begin{equation} C_{m,m}=0, \; D_{mn}=D_{nm},\; D_{mm}=0,\; D_{mn,n}=0,\;
 G_{m,m}=0.
\end{equation}
A special solution of the E-L equations (3.9) is then 
(here $\Delta \equiv \partial_n \partial_n$)

\begin{equation}
 \xi^{(s)}_m ({\bf x},t) = - 
\widetilde{a}^2(t)\Big\{ \frac{1}{2} 
\Delta^{-1} (A-2 \frac{\dot{\widetilde{a}}(t)}{\widetilde{a}(t)}\xi_0),_m 
+ \frac{1}{2} 
B,_m({\bf x},t) + C_m({\bf x},t) 
\Big\}.
\end{equation}
Introducing this into the transformation gives ${h'}_{mn}$ as in (3.10), where
now

\begin{eqnarray}
A' & = & A-2\frac{\dot{\widetilde{a}}(t)}{\widetilde{a}(t)}\xi_0,\;
B' = - \Delta^{-1} ( A-2\frac{\dot{\widetilde{a}}(t)}{\widetilde{a}(t)}\xi_0),
\;{C'}_m=0, \; {D'}_{mn} = D_{mn}, \mbox{ i.e.}
\nonumber\\
h'_{mn} & = & \widetilde{a}^2(t') 
[(\partial _m \partial _n - \delta_{mn} \Delta ) B' + D_{mn}].
\end{eqnarray}
We have $A' + \Delta B' =0$, which implies that ${h'}_{mn}$ is transversal, 
${h'}_{mn},_n = 0$.

The solution 
$\xi^{(s)}_m$  in (3.13) is not unique, 
because neither the operator $\Delta^{-1}$  nor 
the decomposition (3.10) is unique. To see this more explicitly,
let $c$ be a constant and
$\varphi , \psi$ functions with $\Delta \varphi = \Delta \psi = 0$. 
Then one can replace
$A \rightarrow A+c, B \rightarrow B - c x^2/2- \psi,
C_m \rightarrow C_m - \varphi ,_m, 
D_{mn} \rightarrow D_{mn} + 2 \varphi ,_{mn} + \psi ,_{mn}$, 
without changing $h_{mn}$. In particular one can always remove a constant 
$D_{mn} = \overline{D}_{mn}$  
by  putting it into $B,_{mn}$ with $B = \overline{D}_{ij}x^ix^j/2$. 
We assume that this has been performed if necessary.
For a detailed discussion we write
the most general solution of (3.9) as 

\begin{equation}
\xi_m ({\bf x},t)  =  \xi^{(s)}_m ({\bf x},t) 
+  \eta_m({\bf x},t). 
\end{equation}
Introducing  this into  (3.8) 
leads to a variation problem for $\eta_m$:

\begin{equation}
\int  \Big(  
  \eta _m,_n + \eta _n,_m  - \frac{2}{3} \langle \eta _i,_i \rangle \delta_{mn}
+{h'}_{mn} - \frac{1}{3} \langle {h'}_{ii} \rangle \delta _{mn} \Big) ^2 d^3x
= \mbox{ Minimum.}
\end{equation}
A variation in the interior results in the  
homogeneous equations associated with (3.9), i.e. 

\begin{equation}
 \eta_m,_{nn} + \eta_n,_{mn} =0.
\end{equation}
Two tasks have to be done. Firstly one has to consider variations
which also involve changes at the boundary. This will yield a special
solution for $\eta_m({\bf x},t)$ which fulfills the boundary conditions.
Secondly one has to classify the whole set of solutions, in order to
determine the remaining freedom. 

Concerning the first point it will turn out that the solution is
concentrated in a strip along the boundary, 
with an extension of the wave length of the perturbation ${h'}_{mn}$. 
 If the wave length is small compared to 
the extension of the volume it is therefore irrelevant.

As for  the second point we will find that the freedom consists in
rigid translations, rotations, 
and scaling, at this stage still with an arbitrary time dependence.

If these results  appear obvious, one may skip the following lengthy
derivations and proceed directly to (3.24), (3.25) at the end of step 1.
\\

We now treat general variations 
$\eta_m \rightarrow \eta_m + \delta \eta_m$ which also involve 
changes at the boundary. This poses a delicate problem, 
because $\eta_m$ has to fulfill the 
homogeneous E-L equations in the interior. Therefore it would not help 
to consider the boundary terms 
from the partial integration because $\eta_m$ cannot be 
chosen completely free at the boundary.
We therefore proceed in the following way.

We know that $\eta_m$ has to fulfill the homogeneous 
E-L equations. 
Consider therefore a complete set of solutions $\eta_m^{[\alpha]}$ of (3.17), 
expand $\eta_m = \sum _\alpha c_\alpha \eta_m^{[\alpha]}$, and introduce into 
the variation problem (3.16). Differentiation with respect to the 
coefficients $c_\alpha$ leads to the linear system of equations

\begin{eqnarray}
& & I^{[\alpha \beta]}c_\beta + 
\int (\eta^{[\alpha]} _{m,n} + \eta^{[\alpha]}_ {n,m}  
- \frac{2}{3} \langle \eta^{[\alpha]}_{i,i} \rangle \delta_{mn})
({h'}_{mn} - \frac{1}{3} \langle {h'}_{jj} \rangle \delta _{mn}) d^3x = 0,
\mbox{ with}
\nonumber\\
& & I^{[\alpha \beta]}  =  \int (\eta^{[\alpha]} _{m,n} + \eta^{[\alpha]} _{n,m}  
- \frac{2}{3} \langle \eta^{[\alpha]}_{i,i} \rangle \delta_{mn})
(\eta^{[\beta]} _{m,n} + \eta^{[\beta]} _{n,m}  
- \frac{2}{3} \langle \eta^{[\beta]}_{j,j} \rangle \delta_{mn})d^3x.
\end{eqnarray}
We may assume that $I^{[\alpha \beta]}$ is diagonalized, and normalized such 
that it has eigenvalues 0 and $L^3$ only, where $L$ is some length 
introduced for dimensional reasons. Eigenvalues 0 belong to solutions 
of (3.17) which in addition fulfill 

\begin{equation}
\eta^{[\alpha]} _{m,n} + \eta^{[\alpha]} _{n,m}  
- \frac{2}{3} \langle \eta^{[\alpha]}_{i,i} \rangle \delta_{mn}=0.
\end{equation} 
Contributions of this type solve (3.18) trivially for 
arbitrary $c_\alpha$ and may always be added. 
The only solutions of these equations are rigid translations, rotations, 
and scaling, at this stage still with an arbitrary time dependence.
To show this formally, we first observe that (3.19) 
implies 
$\eta^{[\alpha]}_{m,nk} + \eta^{[\alpha]}_{n,mk} = 0$
(at this point it 
becomes apparent why we used the average $\langle g'_{ii}/3 \rangle$ instead
of $g'_{ii}/3$ in (2.3), (3.8)). 
Applying (3.19) again in order to 
exchange $m,k$ 
and $n,k$, respectively, implies $-2 \eta^{[\alpha]}_{k,mn} = 0$, 
i.e. all second 
derivatives vanish. Thus $\eta^{[\alpha]} _m$ can only contain constant 
and linear 
contributions. Introducing a last time into (3.19), one finds that the 
linear part is restricted to a scaling and a rotation.

The coefficients which refer to the non trivial 
solutions associated with the eigenvalues $L^3$ are uniquely fixed, namely

\begin{eqnarray}
c_\alpha & = & - \frac{1}{L^3}\int (\eta^{[\alpha]} _{m,n} + \eta^{[\alpha]}_ {n,m}  
- \frac{2}{3} \langle \eta^{[\alpha]}_{i,i} \rangle \delta_{mn})
({h'}_{mn} - \frac{1}{3} \langle {h'}_{jj} \rangle \delta_{mn}) \, d^3x 
\nonumber\\
& = & - \frac{2}{L^3} \int \eta^{[\alpha]} _m 
({h'}_{mn} - \frac{1}{3} \langle {h'}_{jj} \rangle \delta_{mn}) n_n dA,
\end{eqnarray}
with $n_n$ the normal vector at the boundary. The term 
$\sim \langle \eta^{[\alpha]}_{i,i} \rangle \delta_{mn}$ in the first line 
does not contribute, in the second line we performed a partial 
integration, making use of ${h'}_{mn},_n = 0$. 

We give examples with plane waves in $z-$direction. 
Consider first a gravitational wave ${h'}_{mn} = \bar{h}_{mn} \cos kz$, with 
$\bar{h}_{11} = - \bar{h}_{22} = \bar{h} = const.$  
as the only non vanishing components. This implies ${h'}_{jj}=0$. 
It is convenient to choose the volume as a cylinder 
with radius $\rho _0$ and $0 \le z \le L $. 
Introduce cylindrical coordinates $\rho,\varphi,z$, together with 
the corresponding unit vectors 
$e_m^{(\rho)}=(\cos \varphi,\sin \varphi,0), \; 
e_m^{(\varphi)}=(-\sin \varphi,\cos \varphi,0), \;
e_m^{(z)}=(0,0,1)$.
Because of $h'_{mz} = 0$ the only surface which contributes in 
(3.20) is $\rho = \rho_0$, where $n_n = e^{(\rho)}_n$. 
One has $e^{(\rho)}_m  \bar{h}_{mn} e^{(\rho)}_n = \bar{h} \cos 2\varphi, \;  
e^{(\varphi)}_m  \bar{h}_{mn} e^{(\rho)}_n = - \bar{h} \sin 2\varphi, \;  
e^{(z)}_m  \bar{h}_{mn} e^{(\rho)}_n = 0.$  This implies that one only 
needs to consider solutions $\eta_m^{[\alpha]}$ with a corresponding structure,
such that the surface integral in (3.20) is non vanishing.
Define the vectors in cylindrical coordinates 
$\eta^{\gamma} = \eta_m e_m^{(\gamma)}, \; \gamma = \rho, \varphi,z$. 
Then a basis of relevant  solutions of the free equation (3.17)  
is obtained by an ansatz of the form 

\begin{equation}
\eta^{(\gamma)} = \left( \begin{array}{cc} 
f(\rho) \cos 2\varphi \cos kz\\
g(\rho) \sin 2\varphi \cos kz\\
h(\rho) \cos 2\varphi \sin kz
\end{array} \right),
\end{equation}
where the  index $(\gamma)$ denotes the components in 
$\rho,\varphi,z$-direction. Obviously
there is no need to consider any other contributions of 
$\cos n \varphi , \; \sin n\varphi $. Either they cannot fulfill (3.17), 
or they give a vanishing $c_\alpha$. If $L$ is a multiple of the wave 
length $\lambda = 2\pi /k$, other modes in $kz$ will also not contribute to 
$c_\alpha$.

Introducing into (3.17) leads to three coupled differential equations for 
the components $\rho,\varphi,z$:

\begin{eqnarray}
2 f'' + 2 \frac{f'}{\rho} 
- 6 \frac{f}{\rho^2} - k^2 f 
+ 2 \frac{g'}{\rho} - 6 \frac{g}{\rho^2} 
+  k h' & = & 0, \nonumber\\
g'' + \frac{g'}{\rho} 
- 9 \frac{g}{\rho^2} - k^2 g 
- 2 \frac{f'}{\rho} - 6 \frac{f}{\rho^2} 
- 2 k \frac{h}{\rho} & = & 0,\\ 
h'' + \frac{h'}{\rho} 
- 4 \frac{h}{\rho^2} - 2  k^2 h 
- k [f' + \frac{f}{\rho} 
+  2\frac{g}{\rho}] & = & 0.\nonumber 
\end{eqnarray}
A (non orthonormalized) basis for the regular solutions is 

\begin{equation}
\left( \begin{array}{cc} 
f^{[1]}\\g^{[1]}\\h^{[1]}
\end{array} \right)
= \left( \begin{array}{cc} 
I_1\\-I_1\\-I_2 
\end{array} \right),\:
\left( \begin{array}{cc} 
f^{[2]}\\g^{[2]}\\h^{[2]}
\end{array} \right)
 = \left( \begin{array}{cc} 
I_3\\I_3\\-I_2
\end{array} \right),\:
\left( \begin{array}{cc} 
f^{[3]}\\g^{[3]}\\h^{[3]}
\end{array} \right)
 = \left( \begin{array}{cc} 
2 k\rho I_2\\0\\k \rho [I_3 -3 I_1] 
\end{array} \right),
\end{equation}
with $I_m \equiv I_m(k\rho )$ the modified Bessel functions with argument 
$k\rho$.

There is no need to go into more details, the qualitative behavior 
can be read off immediately. The solutions $\eta_m^{[\alpha]}$ 
will be combinations of modified
Bessel functions $I_m(k\rho )$. These are monotonically increasing and 
behave like $(k\rho/2)^m/m!$ for small 
$k\rho $, and like $\exp(k\rho) /\sqrt{2\pi k\rho}$ for large $k\rho $.
If the radius of the averaging volume is large compared to the wave length, 
i.e. if $k\rho _0 \gg 1$, the modified Bessel functions as well as the shifts 
$\eta^{(\gamma)}$  are only relevant in a small strip along the boundary  
$\rho = \rho_0$, with an 
extension of the order of $\lambda = 2\pi/k$.

One can normalize $I^{[\alpha\beta]}$ by multiplying $\eta^{[\alpha]}_m$ 
in (3.23) by factors $ \sim L \exp(-k\rho_0)$. 
For the orthonormalized solutions one gets  
$c_\alpha\eta^{[\alpha]}_m \sim (\bar{h}/k) 
\sqrt{\rho_0/\rho} \exp(-k \rho_0) \exp(k\rho)$, i.e.   
$c_\alpha\eta^{[\alpha]}_m \sim \bar{h}/k$ at the boundary, independent 
of $\rho_0$ and $L$.
Therefore, if desired, one may perform the limit $\rho_0 \rightarrow \infty$, 
and/or $L \rightarrow \infty$ and one is sure that the solution 
stays finite.
 
If the boundaries for $z$ are less convenient, i.e. if
$L$ is not a multiple of $\lambda$, the 
situation is slightly more complicated. Instead of (3.21) one needs
a superposition of terms with 
$\cos 2 \pi nz/L$ and $\sin 2\pi nz/L,\;n=0,1,2,\cdots $. The coefficients
in front are only sufficiently large for $n$ with $n \sim kL/2\pi$, 
the coefficients for smaller $n$ decrease in the same way as 
the extension of the modified Bessel functions increases, again the 
solution is concentrated at the boundary $\rho = \rho_0$.

In our next example we consider a wave with $B' = \bar{B} \cos kz$,
which implies ${h'}_{mn} = \bar{h}_{mn} \cos kz$, with 
$\bar{h}_{11}  = \bar{h}_{22} = \bar{h} = \bar{B} k^2$ as the only 
non vanishing 
components. Again, for a wave, $\langle {h'}_{jj} \rangle \approx 0$. 
We now have
$e^{(\rho)}_m  \bar{h}_{mn} e^{(\rho)}_n = \bar{h}, \;  
e^{(\varphi)}_m  \bar{h}_{mn} e^{(\rho)}_n =   
e^{(z)}_m  \bar{h}_{mn} e^{(\rho)}_n = 0.$ 
Therefore one can use an ansatz like (3.21), with 
$\cos 2 \varphi$ and $\sin 2 \varphi$ replaced by 1. The further calculation
proceeds as before with an analogous result.

The previous findings are very convenient. For a large volume one can 
neglect the boundary effects, or alternatively, apply the result of the 
transformation only within a slightly smaller volume.

After step 1 we have transformed to a system where

\begin{eqnarray}
g^{(1)}_{mn}  & = & \widetilde{a}^2(t') \delta_{mn} + h'_{mn} 
= \widetilde{a}^2 (t')[\delta_{mn} 
+ (\partial _m \partial _n - \delta_{mn} \Delta ) B' + D'_{mn}] ,
\nonumber\\
& & \mbox{ i.e. } 
A' + \Delta B'=0, \;C'_m=0, \; \Rightarrow {h'_{mn},}_n = 0.
\end{eqnarray}
The freedom
of remaining  transformations in the new system, due to the freedom
in the homogeneous solution $\eta^m$, is now restricted to 
 
\begin{eqnarray}
\xi^m & = & \xi _m/\widetilde{a}^2(t) = b^m(t) + S(t) x^m 
+ [\omega (t) \times x]^m, 
\nonumber\\
\xi_0 & & \mbox{ completely arbitrary}.
\end{eqnarray}
The functions $b^m(t),S(t),\omega^k(t)$ still have an arbitrary 
time dependence at this stage.

Before proceeding we will assume that the above transformations have been 
performed, 
such that (3.24), (3.25) hold. These conditions  have 
to be maintained in the following steps.  We now drop the primes 
for the optimized system obtained after step 1, and use the prime for 
the optimized system of step 2. 
\\

{\bf Step 2:}

According to (2.4) we have to minimize

\begin{equation}  \int \Bigg( \frac{{g'}_{m0}(x')}
{\sqrt{\langle -{g'}_{00} \rangle
 \langle {g'}_{ii}/3 \rangle}}  \Bigg)^2 \sqrt{-g'(x')}d^4x'  =
\int \widetilde{a}(t) \Big( h_{m0} +   \xi _0,_m + \widetilde{a}^2 
\xi^m,_0 \Big) ^2 d^4x,
\end{equation}
where $\xi^m$ is now restricted to the special 
form (3.25). The functions in the minimization procedure are 
$b^m(t),S(t),\omega^k(t)$, and $\xi_0({\bf x},t)$.

Introducing the decomposition (3.11) and writing the terms with
$b^m(t)$ and $S(t)x^m$ in $\xi^m$ as gradients one obtains

\begin{eqnarray}
h'_{m0} & = & h_{m0} +   \xi _0,_m + \widetilde{a}^2(t) \xi^m,_0 
\\
& = & \Big\{ \xi_0 + \widetilde{a}(t)F 
+ \widetilde{a}^2(t) [\dot{b}^k(t)x^k 
+ \dot{S}(t)x^2/2] \Big\} ,_m + \widetilde{a}(t)G_m 
+ \widetilde{a}^2(t) [\dot{\omega}(t) \times x]^m.
\nonumber
\end{eqnarray}
Obviously one cannot determine $\xi_0,\dot{b}^k(t), \dot{S}(t)$ separately
because they only enter in the combination 

\begin{equation}
\tilde{\xi}_0({\bf x},t) = \xi_0({\bf x},t) + \widetilde{a}(t)F 
+ \widetilde{a}^2(t) [\dot{b}^k(t)x^k + \dot{S}(t)x^2/2]. 
\end{equation}
Therefore we may, at this stage, choose 
$\dot{b}^k(t)$ and $\dot{S}(t)$ arbitrary and vary only 
$\tilde{\xi}_0({\bf x},t)$ and $\dot{\omega}^k(t)$.

Only for simplicity and the sake of obtaining more transparent formulae, 
from now on we specialize to the case that the volume is
a sphere of radius $r_0 = r_0(t)$ around the origin. Obviously 
this condition  
is coordinate dependent, nevertheless it does not destroy the covariance of 
the procedure. 
In any other system, obtained by an infinitesimal transformation, 
the volume is a distorted sphere with an infinitesimal modification 
of the boundary. This only introduces irrelevant boundary 
effects of higher order.

The E-L equation and boundary condition 
 ($n_m$ denotes the normal vector at the boundary of the three
 dimensional sphere) for $\tilde{\xi}_0$ become

\begin{eqnarray}
\Delta \tilde{\xi}_0 & = & 0,
\\
( \tilde{\xi}_0,_m + \widetilde{a}(t)G_m ) n_m & = & 0 \mbox{ at the boundary.}
\nonumber
\end{eqnarray}
We dropped the boundary term 
$ \widetilde{a}^2(t) [\dot{\omega}(t) \times x]^mn_m$
which vanishes for a sphere.

The situation is similar to that in the first step, but simpler because 
we now can use the Neumann type boundary conditions for $\tilde{\xi}_0$.
As an example consider a plane transversal wave 
$G_m = \overline{G} \delta_{mx} \cos kz$. This gives 
$G_mn_m = \overline{G} \sin \theta \cos \varphi \cos (kr_0 \cos \theta).$
The solution of (3.29) can then be expanded as 
$\tilde{\xi}_0 = \overline{G} \sum_{l=1}^\infty 
c_l (r/r_0)^l P_l^1(\cos \theta) \cos \varphi ,$
with $c_l = - [\widetilde{a}(t)r_0(2l+1)/2l^2(l+1)] \int_0^\pi 
\sin^2 \Theta P_l^1(\cos \Theta ) \cos (kr_0 \cos \Theta) d\Theta.$ 
If $kr_0 = 2 \pi r_0/\lambda \gg 1$, the factor 
$\cos (kr_0 \cos \Theta)$ oscillates rapidly, therefore
the integral is only important if this oscillation matches with the 
oscillation of the Legendre polynomial, i.e. if $l \approx 2 kr_0 /\pi$. 
This is large, therefore the factor $(r/r_0)^l$ is only relevant 
near the boundary. In the interior the solution is essentially zero. 
We may therefore replace (3.29)
by the corresponding Neumann problem with $G_m \equiv 0$ which has 
the unique solution $\tilde{\xi}_0 = - \tau(t)$, where $\tau$ 
is constant in space but may depend on time.
Therefore (3.28) gives

\begin{equation}
\xi_0({\bf x},t) = - \tau(t) - \widetilde{a}(t)F   
-  \widetilde{a}^2(t) [\dot{b}^k(t) x^k + \dot{S}(t)x^2/2].
\end{equation}
Next we vary $\dot{\omega}^k(t)$. Because this is a function of $t$ only, 
the $x$-integration remains and we obtain

\begin{equation}
\dot{\omega}^k(t)  =  - \frac{3}{2 \widetilde{a}(t) \, \langle x^2 \rangle }
\epsilon_{kln} \langle x^l G_n  \rangle.
\nonumber
\end{equation}
We used some simplifications for the case 
of a sphere, and the result just obtained for $\tilde{\xi}_0$,

After applying the transformation in (3.27) with the results (3.30) and (3.31) 
we have transformed to a system where 

\begin{eqnarray}
h'_{m0} & = & \widetilde{a}(t') \, G'_m \mbox{, with }
G'_m = G_m + 3 \widetilde{a}(t') \, \langle x^m G_n 
- x^n G_m \rangle x^n /2,
\nonumber\\
\mbox{ i.e. } F' & = & 0, \; \Rightarrow {h'}_{m0},_m = 0.
\end{eqnarray}
The remaining freedom for coordinate transformations in 
the new system is now

\begin{eqnarray}
\xi^m & = & \xi_m/ \widetilde{a}^2(t) =
 b^m(t) + S(t)x^m + [\omega \times x]^m ,
\nonumber\\
\xi^0 & = & -\xi_0 = \tau(t) 
+ \widetilde{a}^2(t) [ \dot{b}^k(t) x^k 
+ \dot{S}(t) x^2/2 ].
\end{eqnarray}
We are no longer free to perform different rotations at different 
times, $\omega ^k= const.$, while translations $b^m(t)$, scalings $S(t)$,
as well as $\tau (t)$, can still be time dependent.

Again we assume that the transformations have been performed and that 
the following steps respect (3.32), (3.33). The primes will be dropped.
\\

{\bf Step 3:}

We have to minimize

\begin{equation} \int \Big( {g'}_{00}(x') + 1 \Big)^2
\sqrt{-g'(x')} d^4x'=  \int \widetilde{a}^3(t) \Big( h_{00} 
+   2\xi {_0,_0} \Big) ^2d^4x .
\end{equation}
Using the result (3.33) for $\xi _0$, the bracket becomes

\begin{equation}
h_{00} + 2 \xi _0,_0 = h_{00} 
 - 2\frac{\partial}{\partial t} \Big\{ 
 \tau (t)
+  \widetilde{a}^2 (t)[\dot{b}^k(t)  x^k + \dot{S}(t)  x^2/2] 
\Big\}.
\end{equation}
Variation of $\tau (t), \dot{b}^k(t), \dot{S}(t)$  gives 
the equations (3.36) - (3.38) below, in which we already performed 
the spatial integrations 
where possible. In the derivation we used some simplifications which 
hold for a sphere, i.e. 
$\langle x^m \rangle = 0, \; 
\langle x^kx^l \rangle =\langle x^2 \rangle \delta^{kl}/3.$ For a sphere 
of radius $r_0$ one has $\langle x^2 \rangle = 3 r_0^2/5, \; 
\langle (x^2)^2 \rangle = 3 r_0^4/7$, and, in the denominators below, 
$\langle (x^2)^2 \rangle  - \langle x^2 \rangle ^2 = 12 r_0^4/175$.

\begin{eqnarray}
\langle h_{00} \rangle  - 2  \dot{\tau }(t) 
- \langle x^2 \rangle \frac{\partial}{\partial t}(\widetilde{a}^2(t)\dot{S}(t)) 
 & = & 0,\\ 
\langle x^m h_{00} \rangle
- \frac{2}{3} \langle x^2 \rangle 
\frac{\partial}{\partial t}(\widetilde{a}^2(t)\dot{b}^m(t)) 
 & = & 0,\\
\langle x^2 h_{00} \rangle  
- 2 \langle x^2 \rangle  \dot{\tau }(t) 
- \langle (x^2)^2 \rangle  \frac{\partial}{\partial t}
(\widetilde{a}^2(t)\dot{S}(t)) 
 & = & 0.
\end{eqnarray}
The solutions are

\begin{eqnarray}
 \dot{\tau}(t) & = &  \frac{1}{2}
\frac{\langle (x^2)^2 \rangle \langle h_{00} \rangle 
- \langle x^2 \rangle \langle x^2 h_{00} \rangle }
{\langle (x^2)^2 \rangle - \langle x^2 \rangle^2},
\\
\frac{\partial}{\partial t} \Big( \widetilde{a}^2(t) \dot{b}^m(t) \Big) & = & 
\frac{3}{2} \frac{\langle x^m h_{00} \rangle }{\langle x^2 \rangle },
\\
\frac{\partial}{\partial t}(\widetilde{a}^2(t)\dot{S}(t)) 
& = & \frac{\langle x^2 h_{00} \rangle 
-\langle x^2 \rangle \langle h_{00} \rangle }
{\langle (x^2)^2 \rangle - \langle x^2 \rangle^2}.
\end{eqnarray}
The function $\tau(t)$ is now fixed up to an additive constant $\tau$,
while  $b^m(t)$ and $S(t)$  are fixed up to two integration constants 
$b^m, \; \beta ^m$, and $S, \; \Sigma $, respectively.
 
\begin{equation}
b^m(t) = \widehat{b}^m(t) + b^m + \beta^m  
\int ^t \frac{dt'}{\widetilde{a}^2(t')}, \;
S(t) = \widehat{S}(t) + S + \Sigma \int ^t \frac{dt'}{\widetilde{a}^2(t')}, 
\end{equation} 
with $\widehat{b}^m(t)$ and $\widehat{S}(t)$  obtained 
from integrating (3.40) and (3.41).

After having performed the transformations of step 3 one has 
the equations 
$\langle h'_{00} \rangle =
\langle x^m h'_{00} \rangle =
\langle x^2 h'_{00} \rangle = 0$ . 
The remaining freedom for transformations is now 

\begin{eqnarray}
\xi^m & = & \xi_m/\widetilde{a}^2(t) = 
 b^m + S x^m + [\omega \times x]^m
 + (\beta ^m + \Sigma \, x^m)  \int ^t\frac{dt'}{\widetilde{a}^2(t')},
\nonumber\\
\xi^0 & = & - \xi_0 = \tau + \beta^k x^k + \Sigma \, x^2/2,
\end{eqnarray}
where $b^m,S,\omega^k,\tau,\beta^m,\Sigma$ are all constant.
As before we assume that the transformations 
have been performed and that the form of further transformations is 
restricted to (3.43).
\\

{\bf Step 4:}

Before proceeding with this step
we look at the meaning of the constants $\beta^m$ and $\Sigma$. 
Obviously $\beta^m$ describes an infinitesimal boost (slightly modified 
because $\widetilde{a}^2(t) \ne const.$), while  
$\Sigma$ describes a time dependent scaling.
A transformation with (3.43) gives 
a space dependent contribution 
$ 2 \widetilde{a} \dot{\widetilde{a}} \, [\beta^k x^k + \Sigma \, x^2/2] 
\, \delta_{mn}$ in $g'_{mn}$, which illustrates that boosts 
do not leave the standard form of the R-W metric invariant. This 
contribution is, 
however, suppressed by the factor $\dot{\widetilde{a}}$, 
therefore a minimization along the previous lines appears ineffective. 
We therefore choose condition (2.6) in step 4, and minimize

\begin{eqnarray}
& & 
\int \Bigg(\frac{\partial}{\partial t'} g'_{00}(x') 
\Bigg)^2 \sqrt{-g'(x')}d^4x' \\ 
& = &
\int \widetilde{a}^3(t) \Big( h_{00},_0    
+ \frac{1}{\widetilde{a}^2(t)}(\beta^k + \Sigma x^k) h_{00},_k \Big) ^2d^4x.
\nonumber
\end{eqnarray}
We are left with an ordinary minimization problem for the four 
constants $\beta^k$ and $\Sigma$. Define the integrals

\begin{eqnarray}
\left\{ \begin{array}{l}I\\I_m\\I_{mn} \end{array} \right\} & = &
\int \frac{1}{\widetilde{a}(t)} 
\left\{ \begin{array}{c}x^mx^n\\x^n\\1 \end{array} \right\}
h_{00},_m h_{00},_n d^4 x,
\nonumber\\
\left\{ \begin{array}{l}J\\J_m \end{array} \right\} & = &
\int \widetilde{a}(t) 
\left\{ \begin{array}{c}x^m\\1 \end{array} \right\}
 h_{00},_m h_{00},_0 d^4 x.
\end{eqnarray}
The linear system for $\beta^k$ and $\Sigma$ then becomes
\begin{eqnarray}
I \Sigma + I_n \beta^n & = & -J
\nonumber\\
I_m \Sigma + I_{mn} \beta^n & = & - J_m.
\end{eqnarray}
This can be easily solved after $h_{00}$ is specified. 
Practically it is even simpler. For a large averaging sphere the 
volume integrals are rotation invariant, which implies $I_m = J_m=0$ and
$I_{mn} = \delta_{mn}I_{ii}/3$, leading to $\beta^m=0, \; \Sigma = -J/I$.

In the transformed 
system we will then have $J'=0, \; J'_m=0$.

The nature of step 4 is somewhat different from the steps before. 
While steps 1, 2, 3 only require that $h_{\mu\nu}$ is small, step 4 
requires in addition that time variations are small compared to spatial 
variations.

Our optimization procedure has now come to an end. The integration 
constants $\beta^m, \; \Sigma$ are fixed. 
After step 4 the only 
allowed transformations which remain are those which keep the form of the 
R-W metric invariant.

\begin{eqnarray}
\xi^m & = & \xi_m /\widetilde{a}^2(t) =
 b^m + S x^m + [\omega \times x]^m ,
\nonumber\\
\xi^0 & = & - \xi_0 = \tau ,
\end{eqnarray}
with $b^m,S,\omega^k,\tau$  all constant.
\\

{\bf Step 5:}

In the last step we determine the optimal scale factor $a(t)$. 
In (3.24) we obtained the form 
$g^{(1)}_{mn} = \widetilde{a}^2(t) [(1-\Delta B) \delta_{mn} + B,_{mn} + D_{mn}]$, 
where we denote the first order result for the optimized metric 
by an index ${(1)}$. According to (2.7) we define

\begin{equation}
a^2(t) \equiv \langle g^{(1)}_{ii} /3 \rangle (t) 
= \widetilde{a}^2(t) [1-\frac{2}{3} \langle \Delta B \rangle ].
\end{equation}
Eliminating $\widetilde{a}^2(t)$  gives $g^{(1)}_{mn} = a^2(t) 
[(1 - \Delta B + \frac{2}{3} \langle \Delta B \rangle )\delta_{mn} 
+ B,_{mn} + D_{mn}]$, which can finally be written in the form

\begin{equation}
 g^{(1)}_{mn} = a^2(t) 
[\delta_{mn} 
+ (\partial _m \partial _n - \delta_{mn} \Delta ) B^{(1)} + D^{(1)}_{mn}],
\end{equation}
with $B^{(1)} = B - \langle \Delta B \rangle x^2/6, \; D^{(1)}_{mn} = D_{mn}$. 
One now has $\langle h^{(1)}_{mm} \rangle =
- 2 a^2 \langle \Delta B^{(1)} \rangle = 0$.
\\

The whole procedure is simpler than it appears. Usually one starts already 
with an ansatz for the fluctuations in a ``reasonable'' coordinate 
system. Then steps 2 - 5 may become trivial from simple symmetry arguments, 
i.e. one is already using the optimal coordinates and no further 
transformations are necessary. 
\\

{\bf Summary of the first order transformation}

We found that the transformed  metric 
$g^{(1)}_{\mu\nu}$ after the first order 
optimization has the form

\begin{eqnarray}
g^{(1)}_{mn}  & = & a^2(t) \delta_{mn} + h_{mn} 
= a^2 (t)[\delta_{mn} + (\partial _m \partial _n - \delta_{mn} \Delta ) B 
+ D_{mn}] ,
\\
g^{(1)}_{m0}  & = & h_{m0} = a(t) G_m,\\
g^{(1)}_{00} & = & -1 + h_{00}.
\end{eqnarray}
The approximating R-W metric $\overline{g}_{\mu\nu}$ is obtained by dropping 
the perturbations $B,D_{mn},$ $G_m,h_{00}$. This provides a
natural basis for splitting into background and perturbation, and 
for performing perturbative calculations.

The quantities which could not be removed by gauge transformations 
nevertheless share some properties of the unperturbed metric.
Several relations can be derived by appropriate partial integrations and
some simplifications which hold for a sphere:

\begin{eqnarray}
h_{mn},_n & = & 0, \; \langle h_{ii} \rangle =  0, \;
\langle x^n h_{mn} \rangle =  0, \;
\langle (x^m x^n + x^2\delta^{mn}/2) h_{mn} \rangle =  0, 
\\
h_{m0},_m & = & 0, \; 
\langle x^m h_{m0} \rangle = 0, \; 
\langle x_m h_{n0} - x_n h_{m0} \rangle = 0,
\\
\langle h_{00} \rangle & = & 0, \; \langle x^m h_{00} \rangle = 0, 
\; \langle x^2 h_{00} \rangle = 0,
\end{eqnarray}
\begin{equation}
\int a^3(t) 
\left\{ \begin{array}{c}x^m\\1 \end{array} \right\}
h_{00},_m h_{00},_0  d^4 x = 0.
\end{equation}
The relations 
$\langle h_{mm} \rangle =  0, \;
\langle h_{00} \rangle  =  0, $ will lead to the absence of 
backreaction in first order.

\setcounter{section}{3}
\setcounter{equation}{0}\addtocounter{saveeqn}{1}%

\section{Second order}

We will see in sect. 5 that the second order of the 
transformation to the optimal gauge is not needed if one neglects time 
derivatives in the background metric and in the perturbed density. 
Nevertheless, from a principle 
point of view, it is instructive to show how the procedure can be 
extended to second order in a quite simple way.

Rather than trying an approach by brute force, one should proceed in an 
iterative way. Determine the shift $\xi _\mu$ which leads to 
the optimal system in first order. If one uses, instead of (3.2) - (3.5), 
the exact transformation formulae, or at least, considers $\xi _\mu$ 
up to second order, one obtains a perturbation of the form
$h_{\mu\nu} = h^{(1)}_{\mu\nu} + \tilde{h}_{\mu\nu}$ in the optimal system. 
All observers, irrespective of their original gauge, have ended up with 
the same $h^{(1)}_{\mu\nu}$, up to the remaining 8-parameter group of rigid 
transformations.
For the higher order contributions 
$\tilde{h}_{\mu\nu}$ which have not been optimized this is not the case. Unlike 
$h^{(1)}_{\mu\nu}$ they do depend on the original gauge. An explicit calculation 
would be tedious.  We will, however, see that $\tilde{h}_{\mu\nu}$ does not 
contribute to backreaction, therefore there is no need to calculate it.

We now start the second order calculation with the metric 
$h_{\mu\nu} = h^{(1)}_{\mu\nu} + \tilde{h}_{\mu\nu}$.
Within the brackets in (2.3) - (2.6) we have to expand the metric 
${g'}_{\mu\nu}$ up to second order in the perturbation and to 
perform the minimizations in the four steps. There is a 
considerable simplification due to the iterative procedure.
Because $h^{(1)}_{\mu\nu}$ is the solution of the  first order problem there 
are no first order terms in $\xi_\mu$.
Consequently $\xi_\mu$ needs only 
to be considered in lowest order, i.e. one can again apply the simple 
transformation formulae (3.2) - (3.5). 

It is convenient to include the factor 
$\sqrt{-g'}$ into the squares by writing 

\begin{equation}
\sqrt{-g'} = a^3 \Big[ 1 + \frac{1}{4}(\frac{h^{(1)}_{ii}}{a^2} -
h^{(1)}_{00}) \Big] ^2.
\end{equation}
The scale factor $\widetilde{a}$ has  been replaced by the optimized 
scale factor 
$a$ of the first order.
The brackets have now to be considered in second order of the perturbation.
Due to the properties 
$\langle h^{(1)}_{ii} \rangle = 0$ and $\langle h^{(1)}_{00} \rangle = 0$,
 there are no corrections from the averages in the  
denominators of steps 1,2.

The shift $\xi_\mu$ is a superposition 
of two terms, $\xi_\mu = \tilde{\xi}_\mu + \xi^{(Q)}_\mu $. The first term, 
$\tilde{\xi}_\mu$, corresponds to the solution of the E-L equations 
 with $h_{\mu\nu}$ replaced by $\tilde{h}_{\mu\nu}$.
This part transforms $\tilde{h}_{\mu\nu}$ into $\tilde{h}^{(2)}_{\mu\nu}$
in the same way as previously it transformed 
$h_{\mu\nu}$ into $h^{(1)}_{\mu\nu}$. Therefore 
$\tilde{h}^{(2)}_{\mu\nu}$ shares the 
properties (3.53), (3.55), i.e. $\langle \tilde{h}^{(2)}_{ii} \rangle = 0$, 
$\langle \tilde{h}^{(2)}_{00} \rangle = 0$, consequently it will not lead 
to any backreaction, and one does not need 
$\tilde{\xi}^{(2)}_\mu$ and $\tilde{h}^{(2)}_{\mu\nu}$ .

The second contribution, $\xi^{(Q)}_\mu$, is due to the quadratic terms in 
$h^{(1)}_{\mu\nu}$ in the brackets in (2.3) - (2.6), which now have to be
inserted into the E-L equations. This part could, in principal, lead to a 
backreaction. As mentioned before, we will however see that this does 
not happen if we drop the time dependence.

\setcounter{section}{4}
\setcounter{equation}{0}\addtocounter{saveeqn}{1}%
 
\section{Averaging}

In first order of the perturbation we finally obtained an optimal 
coordinate system with a metric of the special form (3.50) - (3.52), 
and the approximating R-W metric $\overline{g}_{\mu\nu}$  obtained by dropping 
the perturbations.
One can now simply define 
spatial averages of arbitrary tensors in the naive 
way as in (2.2), i.e. 

\begin{equation}
\langle {A}_{\mu\nu\cdots} \rangle (t) = 
\frac{\int {A}_{\mu\nu\cdots}({\bf y},t) \sqrt{^3g({\bf y},t)} d^3y}
{\int \sqrt{^3g({\bf y},t)} d^3y}.
\end{equation}
The freedom in the choice of coordinates is restricted to the eight 
parameter group of global transformations described by the parameters
$b_m,S,\omega_k,\tau$, and these transformations commute with the 
operation (5.1) of averaging.

It is not necessary to restrict to averaging 
of scalar quantities or to decompose into invariants
(usually only with respect to purely spatial transformations)
and perform the averages for the latter..
In fact such an approach can be problematic. The expansion tensor 
$\Theta_{mn} = g_{mn},_0/2 $ is often decomposed into invariants, 
and quantities which enter linearly and quadratically are averaged 
separately (see e.g \cite{Buchert}). This procedure can lead to quite 
strange consequences
like negative averages of positive definite expressions. Nothing of this 
kind can happen in our case.

For the special case of the metric we note that 
 $\langle g_{\mu\nu} \rangle$  is not necessarily 
identical with $\overline{g}_{\mu\nu}$. Due to (3.53), (3.55) 
the relation is, however, true 
for  $g_{00}$ and for  $g_{mm}$ in first order, i.e. 
$\langle g^{(1)}_{00} \rangle = \overline{g}_{00}$
and   $\langle g^{(1)}_{mm} \rangle = \overline{g}_{mm}$.
These are the relevant quantities for 
backreaction which we discuss now. 

\setcounter{section}{5}
\setcounter{equation}{0}\addtocounter{saveeqn}{1}%

\section{Backreaction}

In the same way as the perturbed metric,  
$g_{\mu\nu} = \overline{g}_{\mu\nu} + \delta g_{\mu\nu}$,
we split the Einstein tensor,  
$G_{\mu\nu} = \overline{G}_{\mu\nu} + \delta G_{\mu\nu}$, 
with $\overline{G}_{\mu\nu}$ 
the Einstein tensor associated with $\bar{g}_{\mu\nu}$. Using the 
Einstein equations (we include a cosmological constant $\Lambda$)
for $G_{\mu\nu}$ one has

\begin{equation}
\overline{G}_{\mu\nu}   =   \langle \overline{G}_{\mu\nu} \rangle 
= \langle G_{\mu\nu} \rangle - \langle \delta G_{\mu\nu} \rangle
=  \kappa \langle T_{\mu\nu} \rangle - \Lambda \langle g_{\mu\nu} \rangle
-\langle \delta G_{\mu\nu} \rangle .
\end{equation}
The first two terms, $\kappa \langle T_{\mu\nu} \rangle
- \Lambda \langle g_{\mu\nu} \rangle $, 
describe the equations which one would expect 
from the averaged energy-momentum tensor and metric, the third one, 
$ -\langle \delta G_{\mu\nu} \rangle $, is the 
deviation from this, i.e. the backreaction. The essential quantities 
associated with density and pressure mimicked by backreaction are 
$\kappa \rho _b= - \langle \delta G_{00} \rangle$ and 
$\kappa p_b = - \langle \delta G_m^m \rangle/3 $. 
In first order of the perturbation one has (indices are raised and lowered
with the background metric $\overline{g}_{\mu\nu}$)

\begin{eqnarray}
\delta G^{(1)}_{00} & = & \frac{1}{2} (h^{ij},_{ij} - h_i^i,_j^j)
- 2 \frac{\dot{a}^2}{a^2} h_i^i + \frac{\dot{a}}{a} h_i^i,_0 
- 2 \frac{\dot{a}}{a} h^i_0,_i,
\\
\delta G^{(1)m}_m & = & \frac{1}{2} (h_i^i,_j^j - h^{ij},_{ij}) 
- \frac{\dot{a}^2}{a^2} h_i^i + \frac{\dot{a}}{a} h_i^i,_0 - h_i^i,_{00} 
+ 2 \frac{\dot{a}}{a} h^i_0,_i
+ 2  h^i_0,_{i0}  
\nonumber\\
& & - 3 ( \frac{\dot{a}^2}{a^2}   + 2 \frac{\ddot{a}}{a})h_{00} 
- h_{00},_i^i - 3 \frac{\dot{a}}{a} h_{00},_0.
\end{eqnarray}
If one inserts $h^{(1)}_{\mu\nu}$ and uses the properties (3.53) - (3.55)
one observes that there are three types of terms in 
 $\delta G^{(1)}_{00}$  and $\delta G^{(1)m}_m$: 
\\

a) terms which vanish,

b) terms where the spatial average vanishes, 

c) terms which are spatial derivatives and can be written as  
surface contributions in the integral which are irrelevant for large volumes.

Therefore $\langle \delta G^{(1)}_{00} \rangle 
= \langle \delta G^{(1)m}_m\rangle = 0$, there is no backreaction in 
first order. This is in fact a property which one expects from any 
reasonable lowest order averaging prescription, where positive and negative 
contributions of the fluctuations should cancel.
\\

We now come to the second order. Here one has two types of contributions.
The first one arises from introducing the second order correction 
$h^{(2)}_{\mu\nu}$ of the metric into the first order correction 
$\langle \delta G^{(1)}_{\mu\nu} \rangle$ of the Einstein tensor. In 
general this does not vanish because $h^{(2)}_{\mu\nu}$ does not 
fulfill the properties (3.53) - (3.55). For an estimate one can, however, 
use a simple static approximation (we will comment on retardation 
below) which is appropriate for the present day universe.
The peculiar velocities of galaxies are small, 
with  $v/c$ of the order of $10^{-3}$. We also do not have sizeable 
perturbations with extremely short wave length which could contribute large 
derivatives. Let the extension in time of the averaging 
volume be small compared to the Hubble time. 
Then the variation of $h_{\mu\nu}$ in time, caused by the slow motion of 
matter, as well as the weak time dependence of $a(t)$, can be neglected in
comparison with the variation in space. It is convenient to fix $a(t_0) = 1$ 
for the present time $t_0$. The surviving parts of 
$\langle \delta G^{(1)}_{00} \rangle$ and  
$\langle \delta G^{(1)m}_m \rangle $ in (6.2), (6.3) only contain 
spatial derivatives,
i.e. the averages vanish up to irrelevant boundary terms. This 
holds irrespective of the special form of $h^{(2)}_{\mu\nu}$ and is,
of course, very convenient. There is no need to calculate $h^{(2)}_{\mu\nu}$.

The second contribution arises from introducing the first order correction 
$h^{(1)}_{\mu\nu}$ of the metric into the second order correction 
$\langle \delta G^{(2)}_{\mu\nu} \rangle $ of the Einstein tensor. 
The rather lengthy expression for $ \delta G^{(2)}_{\mu\nu}$ can be
found in Wetterich \cite{Wetterich} in the approximation that derivatives
acting on the background metric are neglected.
It will not be written down here. 
We only give the two quantities needed for the effective density and pressure,
anticipating $h_{m0}=0$, neglecting the time dependence 
in the perturbation, and performing some spatial partial integrations.

\begin{eqnarray}
\langle \delta G^{(2)}_{00} \rangle & = & \langle \frac{1}{2}h_{00} h_i^i,_j^j 
- \frac{1}{2} h_{00} h^{ij},_{ij} 
+ \frac{1}{8} h^i_i  h^j_j,_k^k + \frac{1}{8} h^{ij} h_{ij},_k^k
-\frac{1}{4} h^{ij} h^k_j,_{ki} \rangle,
\\
\langle \delta G^{(2)m}_m \rangle & = & \langle -\frac{1}{2} h_{00} h_{00},_i^i
+ \frac{1}{2} h_{00} h^{ij},_{ij} + \frac{3}{8} h^i_i h^j_j,_k^k
- \frac{5}{8} h^{ij} h_{ij},_k^k 
 -  h^i_i h^{jk},_{jk} + \frac{5}{4} h^{ij} h^k_j,_{ki}\rangle  
\end{eqnarray}
In order to estimate $h_{\mu\nu} \equiv h^{(1)}_{\mu\nu}$  from the sources 
we use again 
the simple  approximation above, i.e. neglect any time dependence.
This is essentially the model  
of Wetterich \cite{Wetterich}. It has the advantage 
that it is transparent and leads to explicit formulae. 

Consider a dust universe with $\rho$ only weakly time dependent and $p=0$. 
The solution for the perturbed metric is only needed in lowest order and 
most conveniently first derived in 
the harmonic gauge (marked by a hat)

\begin{equation}
\hat{h}^\nu_\mu,_\nu = \frac{1}{2} \hat{h}^\nu_\nu,_\mu,
\end{equation}
where the perturbed Friedmann equations have the simple form

\begin{equation}
\hat{h}_{\mu\nu},_\rho^\rho = - 2 \kappa
(\delta \hat{T}_{\mu\nu} - 
\frac{1}{2} \delta \hat{T}^\rho _\rho \; \overline{g}_{\mu\nu}). 
\end{equation}
In our simple model $\delta \hat{T}_{00} = \delta \rho$ is  
time independent while all the other components of 
$\delta \hat{T}_{\mu\nu}$  vanish. This implies

\begin{equation}
\Delta \hat{h}_{00} = -\kappa \delta \rho \mbox{ , furthermore }
\hat{h}_m^n = \delta_m^n \hat{h}_{00},\; , \;\hat{h}_{m0} = 0.
\end{equation}
One can transform back from the harmonic gauge to our optimal 
(transversal) gauge by a purely spatial shift
$\xi ^m = - \Delta^{-1} \hat{h}_{00},_m/2$ which leads to

\begin{equation}
h_{00}= \hat{h}_{00}, \; 
h_m^n = (\delta_m^n - \Delta^{-1} \partial_m \partial^n)h_{00}, \; h_{m0} = 0,
\; \delta T_{\mu\nu} = \delta \hat{T}_{\mu\nu}.
\end{equation}
Introducing (6.9) into (6.4), (6.5), all terms can be expressed by 
$\langle h_{00} \Delta h_{00} \rangle 
= - \kappa \langle h_{00} \delta \rho \rangle $, and one obtains

\begin{eqnarray}
\rho_b & = & -\frac{1}{\kappa} \langle \delta G^{(2)}_{00} \rangle 
  =   \frac{7}{4} \langle h_{00} \delta \rho \rangle
\\
p_b & = &
- \frac{1}{3 \kappa }\langle \delta G^{(2)m}_m \rangle 
 =   - \frac{1}{12}  \langle h_{00} \delta \rho \rangle 
= - \frac{1}{21} \rho_b.
\end{eqnarray}
The pressure term coincides with the result of \cite{Wetterich}, the
density term has a factor 7/4 instead of 9/4 in \cite{Wetterich}. This shows
that the correction is smaller in our ``optimal'' gauge as 
one would expect. 
It also illustrates that differences between reasonable gauges are small.

We now consider a more specific simple model. We start with a hierarchy
of clusters, composed of galaxies, composed of dark matter and stars.
Subsequently we will also discuss the modifications due to the 
presence of voids.

Consider first a space filled with homogeneous spherical objects
of radius $L$, density $\hat{\rho}$, and mass $m = 4\pi \hat{\rho} L^3/3$, 
which are roughly
 uniformly distributed in space at positions ${\bf r}_i$. 
The average density is denoted by $\overline{\rho}$.
The corresponding density fluctuation is 

\begin{equation}
\delta \rho ({\bf r}) \equiv \rho ({\bf r}) - \overline{\rho} 
= \hat{\rho } \sum_i \Theta(L-|{\bf r}-{\bf r}_i|) - \overline{\rho},
\end{equation}
such that the average $\overline{\delta\rho}$ vanishes.
To find a useful approximation to the corresponding $h_{00}({\bf r})$ 
we first define a 
distance $D$ by the requirement that the average of $\delta \rho$ vanishes
within a sphere of radius $D$ around a source, i.e. 
$D^3/L^3 = \hat{\rho}/\overline{\rho}$.
This $D$ is roughly half of the average distance between the spheres. 
For well separated sources one has $D \gg L$.
A solution of (6.8) within this sphere (chosen around 
the origin for simplicity) is then

\begin{equation}
h_{00}({\bf r}) = \kappa \Big\{ \hat{\rho }\:
(\frac{L^2}{2} - \frac{r^2}{6})\Theta(L-r)
+ \hat{\rho }\, \frac{L^3}{3r} \Theta(r-L) + 
\overline{\rho} \: \frac{r^2}{6} + \hat{\rho }\, c \Big\}.
\end{equation}
This is just the well known potential of a uniformly charged 
sphere in a constant background. The constant $c$ has been introduced 
in order to achieve $\int_{r \le D} h_{00} d^3x=0$. It is of the order  
$L^3/D$ and will turn out to be irrelevant. 
Of course (6.13) is not an exact solution 
of (6.8), (6.12). The spheres of 
radius $D$ around the sources overlap in some areas and leave 
empty regions elsewhere. Furthermore 
$h_{00}  \sim \kappa \hat{\rho }L^2/3 \ne 0$ at
the boundary of the circle, thus introducing boundary contributions there. 
The average of the fictitious mass distribution associated with
these corrections vanishes, and the location is a distance 
$\approx D \gg L$ away 
from the sources where $h_{00}$ enters in (6.10), (6.11). Therefore the
resulting corrections to $h_{00}$ are suppressed compared to (6.13) and 
can be ignored. The average $\langle h_{00} \rangle$ will already vanish 
when taken over regions involving only a modest number of 
sources. This implies that also $\langle x^m h_{00} \rangle = 0$ and 
$\langle x^2 h_{00} \rangle = 0$, as required in (3.55).

We are interested in the ratio 
$\rho_b/\overline{\rho} = (7/4) 
\langle h_{00} \; \delta \rho \rangle /\overline{\rho}$.
When calculating $\langle h_{00} \delta \rho \rangle$ one can drop 
the constant $- \overline{\rho}$ in (6.12) because 
$\langle h_{00} \rangle = 0$, i.e. one can replace 
$\delta \rho ({\bf r})$ by $\rho ({\bf r})$. 
Only the regions $|{\bf r}-{\bf r}_i| \le L$ where $\rho({\bf r}) \ne 0$ 
contribute in the product.
This results in a considerable simplification.
It is now straightforward to calculate $\langle h_{00} \delta \rho \rangle $ 
from (6.12) and (6.13).
Dropping corrections which are suppressed by higher powers of $L/D$
one finds the following result which can be written 
in various useful ways.

\begin{equation}
\rho_b/\overline{\rho} = \frac{7}{4} \langle h_{00} \, 
\delta \rho \rangle /\overline{\rho}
= \frac{7}{10} \kappa \hat{\rho }L^2 
=  \frac{7}{10} \kappa \overline{\rho} \; \frac{D^3}{L} 
= \frac{21}{40 \pi} \kappa \frac{m}{L} 
= \frac{21}{10} h_{00}(L).
\end{equation}
\\
Before proceeding we give a more careful justification for our neglect of 
time dependence.
It has been argued by Kolb, Matarrese, Notari, Riotto \cite{Kolb}, 
as well as by Bochner \cite{Bochner}, that retardation effects,
though irrelevant for nearby sources, may become important 
when taking into account contributions from distant regions. 
We thus look at retardation effects. 
Let $D$ be the average distance between the sources, $v \ll c$ their 
average velocity, and define a distance $\overline{D}$ such that 
$(v/c) D \ll \overline{D} \ll D$. Retardation is negligible as long 
as the sources can only move a small fraction of $D$ within the
retardation  time,
i.e. for $(v/c) |{\bf r} - {\bf r}'| \le \overline{D} \ll D$. In the 
additional contribution $\tilde{h}_{00}$, where retardation might 
become relevant, one has the retarded solution  
$\tilde{h}_{00}({\bf r},t) = (\kappa/4\pi)
\int \delta \rho ({\bf r} + {\bf r}'',t-r''/c) d^3x''/r''$,
where the integration is only over the region 
$r'' \equiv |{\bf r} - {\bf r}'| \ge (c/v) \overline{D} \gg D$.
Because $r''$ is large compared to the distance of the sources, and 
$\delta \rho $  homogeneous on average,  this expression is independent
of ${\bf r}$, i.e. it would contribute a 
non trivial, time dependent but space independent
$\tilde{h}_{00}$. But this is impossible in our gauge
because it would contradict the condition $\langle h_{00} \rangle = 0$.

Here it becomes clear that the question of  
backreaction is not only a problem of using a reliable approximation for 
the distribution of matter and a consistent mathematical treatment. 
It is also crucial to connect observations 
with statements about the metric, in particular about the scale factor 
which describes the expansion. If observations refer to something like
our ``optimal gauge'' there is no back reaction from retardation. If,
on the other hand, they refer to some ``bad'' gauge in the past, 
there might be backreaction effects. Observations use light which 
essentially moves through a space which expands according to the
average density $\overline{\rho}$,
the corresponding average fluctuation $\overline{\delta \rho}$ 
vanishes. Therefore we don't expect a sizable additional 
contribution $\tilde{h}_{00}$.
\\

If the universe would be only built up from clusters, and if clusters 
would be homogeneous objects, (6.14) would be the final result and 
one should use it with 
the assignment $\hat{\rho }\rightarrow \rho_C, \; L \rightarrow L_C, \; 
D \rightarrow D_C, \; m \rightarrow m_C$. 
We keep this contribution and next consider the additional effect 
that clusters are made up from galaxies. Because the average effect of
clusters has already been considered, the additional density fluctuation 
within a cluster is described by an expression like (6.12),
where now ${\bf r}_i$ denotes the position of galaxies, $\hat{\rho }$ 
has to be taken as the density within a galaxy, and $\overline{\rho}$
is replaced by the average density $\rho_C$ of a cluster. 
The product of $h_{00}$ 
and $\delta \rho$ in $\langle h_{00} \delta \rho \rangle $ 
does not contain mixed terms between the expressions for 
clusters and for galaxies. The reason is that the 
terms referring to galaxies contain contributions which are 
located at distances $\approx D_G$ which are small compared to the extension 
$L_C$ of the clusters. Therefore one may replace them 
by their averages over the cluster which vanishes.
Within one cluster we thus obtain again the contribution (6.14) 
where now the parameters are those for galaxies, 
and the average as well as the average density refer to a 
single cluster. Because of $\overline{\rho} = \rho_C V_C/V_{tot}$
one has 
$\langle h_{00} \; \delta \rho \rangle _C/ \rho_C
= \langle h_{00} \; \delta \rho \rangle / \overline{\rho}$.
If only a fraction $\eta $ of galaxies is located in clusters and the 
fraction $1-\eta $ roughly uniformly distributed, the expression for
the clusters is multiplied by a factor $\eta $, while $\eta $ 
cancels in the two contributions for galaxies.

Finally consider that galaxies are made up of stars and dark matter. 
Because only a fraction of matter is made up of stars, while dark matter is 
assumed uniformly distributed, there is a factor $\Omega_b/\Omega_m$. 
Thus the contributions of clusters, galaxies, and stars add up to

\begin{eqnarray}
\rho_b/\overline{\rho}
& \approx & \frac{21}{40 \pi} \kappa \Big\{
\eta \frac{m_C}{L_C} +\frac{m_G}{L_G} 
+ \frac{\Omega_b}{\Omega_m} \frac{m_S}{L_S} \Big\},
\\
w_b & \equiv & p_b/\rho_b =- \frac{1}{21}.\end{eqnarray}
This expression has the same structure as
 the result in \cite{Wetterich}, but with the definite factor 
$21/40 \pi = 0.17$ in front, which, of course, should only be 
considered as a rough estimate. 
Our ratio $w_f = p_f/\rho_f = -1/21$ is slightly different from the ratio 
$-1/27$ obtained  in \cite{Wetterich}, which is due to the different 
gauges. 

Galaxy clusters and galaxies are not distributed uniformly, but there 
exist large voids, surrounded by bubble walls. Because this fact
plays some role in the discussion on backreaction, let us 
discuss  the implications. Again we use a simple model which
shows the essential features and can be treated explicitly. We describe the 
bubble wall by a uniform distribution of matter in a shell of radius $R$, 
thickness $2L$,  density $\hat{\rho }$, and total mass $m$. 
For $L\ll R$, which we assume, one has $\hat{\rho } = m/8 \pi R^2 L$. Let the 
bubbles lie close together 
with only little space left in between. The density fluctuation
for a bubble centered around the origin is then

\begin{equation}
\delta \rho(r) = \hat{\rho } \; \Theta(L-|r-R|) - \overline{\rho},
\end{equation}
with $\overline{\rho} = \hat{\rho } \; 6L/R$. The solution of (6.8)
with $\langle h_{00} \rangle = 0$ becomes

\begin{eqnarray}
h_{00}(r) = \kappa \hat{\rho } \Big\{ \hspace{-1ex} & - & 
\frac{3}{5} RL \Theta(R-L-r)
\nonumber\\
& + &(\frac{R^2+L^2}{2} - \frac{8}{5} RL  - \frac{r^2}{6} - \frac{(R-L)^3}{3r}) 
\Theta(L-|r-R|) 
\nonumber\\ 
& + & ( \frac{2R^2 L + 2 L^3/3}{r} - \frac{13}{5} RL )\Theta(r-R-L) 
+ \frac{L}{R} r^2 +c \Big\},
\end{eqnarray}
with $c=O(L^2)$. 
In contrast to the case of matter located inside a sphere, one can here
perform the limit $L \rightarrow 0$, while keeping the total mass fixed. 
This results in

\begin{equation}
h_{00}(r) = \frac{\kappa m}{8 \pi R} 
\Big\{ \hspace{-1ex}  -  \frac{3}{5}  \Theta(R-r)
 +  ( 2 \frac{R}{r} - \frac{13}{5} )\Theta(r-R)  + \frac{r^2}{R^2} \Big\}.
\end{equation}

Following the same steps as before one arrives at 
the back reaction due to bubble walls,

\begin{equation}
(\rho_b/\overline{\rho})_{BW} = \frac{7}{4} \langle h_{00} \, 
\delta \rho \rangle /\overline{\rho}
= \frac{7}{10} \kappa \hat{\rho } R L 
=  \frac{7}{60} \kappa \overline{\rho} \; R^2 
=\frac{7}{80 \pi} \kappa \frac{m}{R}= \frac{7}{4} h_{00}(R).
\end{equation}

We next estimate the various contributions by inserting some
standard values.

For clusters we use $m_C = 10^{15} m_\odot = 2 \cdot 10^{48}g$, and 
$L_C = 5 Mpc = 1.5 \cdot 10^{25} cm$.
For galaxies we insert the values for the milky way, 
$m_G = 10^{12} m_\odot = 2 \cdot 10^{45} g$, and  
$L_G =  100 kpc = 3 \cdot 10^{23} cm $, where both numbers include, 
of course, both baryonic and dark matter. 
For the contribution of the stars, finally, we insert the values of the sun, 
$m_S = m_\odot = 2 \cdot 10^{33} g, \; L_S = L_\odot = 7 \cdot 10^{10} cm$. 

For the ratios mass over radius one obtains

\begin{eqnarray}
m_C/L_C & = & 1.3 \cdot 10^{23}g \, cm^{-1} 
\\ 
m_G/L_G & = & 0.7  \cdot 10^{22}  g \, cm^{-1} 
\\
m_S/L_S & = & 3 \cdot 10^{22} g \, cm^{-1}. 
\end{eqnarray}
We further use 
$\kappa  =  2 \cdot 10^{-27} g^{-1} cm$ and  
$\Omega_b/\Omega_m  = 0.17$. For all three cases one has 
$|h_{00}(r)| \le h_{00}(0) < 0.3 \cdot 10^{-4}$, furthermore
$h_{00}$ is smoother than $\delta \rho$. Serious doubts 
concerning the validity of perturbation theory appear inappropriate.
Nevertheless one can often read the argument that a perturbative 
expansion would be inadmissible because $\delta \rho /\overline{\rho}$
is large. One should take a closer look at this ``argument''. In our
model 
$\delta \rho_{max} = 3m/4\pi L^3, \; \overline{\rho} \approx 3m/4\pi D^3$,
therefore indeed $\delta \rho_{max}/\overline{\rho} \approx D^3/L^3$
is large. But there is no reason to panic, because this quantity 
has nothing to do with the perturbation expansion. The relevant dimensionless
quantity which enters is not $\delta \rho /\overline{\rho}$, but
$\kappa m/L$ which is small, less than $3 \cdot 10^{-4}$ in all cases! 

The three contributions in (6.15) become

\begin{eqnarray}
(\rho_b/\overline{\rho})_C & =  & \eta \cdot 0.4 \cdot 10^{-4}
\\
(\rho_b/\overline{\rho})_G & = &  2 \cdot 10^{-6}
\\
(\rho_b/\overline{\rho})_S & = & 1.5 \cdot 10^{-6}.
\end{eqnarray}
All these corrections due to backreaction are very small.

Let us look at some other stellar objects with a larger ratio $m/L$ 
which might be relevant.

For white dwarfs  the ratio $m/L$ is about a factor of $30 \div 100 $ 
larger than that of the sun, the number of white dwarfs is
estimated as about $10\%$ of all stars. Therefore their contribution is
also unimportant.

For neutron stars and black holes one has $m/L = 6 \cdot 10^{27} g \; cm^{-1}$,
which is a factor of 
$2 \cdot 10^5$  larger than the ratio for the sun. 
If $\eta_{NS}$ and $\eta_{BH}$ denote the fraction of neutron 
stars and black holes one obtains a contribution 
$(\rho_b/\overline{\rho})_{NS+BH} = (\eta_{NS}+\eta_{BH})\cdot 0.3$.
This is still a moderate contribution, but the number 
of these objects becomes relevant. Neutron stars are 
supposed to provide a portion of less than $1\%$
of all stars, one expects a contribution not larger than $10^{-2}$. 
Unless there is an extremely large number of black holes 
the latter will also only give a small contribution. 

Finally let us look at the implications of voids, 
surrounded by bubble walls. The average density of matter is 
$\overline{\rho} = 2 \cdot 10^{-30} g \; cm^{-3}$, 
for the radius of the voids we take $R = 20 Mpc = 6 \cdot 10^{25} cm$. 
From (6.20) we then obtain 
$(\rho_b/\overline{\rho})_{BW} =  2 \cdot 10^{-6}$.
The smallness of this contribution is surprising at first sight, 
therefore it deserves a comment. The result can be
easily understood. The distribution of matter in the bubble wall is 
homogeneous in the two tangential directions. Only in radial 
direction it is concentrated in a small shell of thickness $2L$. 
The situation is therefore
essentially one-dimensional. But while in three or two dimensions the Green
function of the Laplacian is $\sim 1/r$ and $\sim \ln r$ 
respectively, i.e. singular at the origin, in one dimension it is 
$\sim r$, which is finite. Therefore one can perform the limit 
$L \rightarrow 0$ in the latter case. No problems arise if one squeezes 
the matter in the surface $r=R$ by taking 
$\rho (r) = m \, \delta(r-R)/4\pi R^2$. The solution $h_{00}(r)$ in 
(6.19) is finite at $r=R$, it just has a kink there. The absence of any 
singular behavior for small $L$
for matter concentrated in walls explains why the effect is so small.

We are aware that our result is in striking contrast to the
statements of  Wiltshire \cite{Wiltshire} who claims that 
the presence of large voids  should lead to a considerable backreaction.
We don't feel in a position to comment on this work. But we doubt whether
a separate  treatment of the metric within and outside the voids is 
appropriate or even legitimate.

The relative unimportance 
of backreaction is, of course, by no means 
a new result. It was already found 
(in a non covariant calculation) by Nelson \cite{Nelson} in 1972, also by
Wetterich \cite{Wetterich} who's model we essentially used, 
as well as by many others.
But certainly the discussion on the (un)importance of back reaction will 
not end in the near future.

\setcounter{section}{6}
\setcounter{equation}{0}\addtocounter{saveeqn}{1}%

\section{Summary and conclusions}

If one wants to go beyond the simple static approximation which we 
used for the calculation of backreaction  one is faced with the problem 
that the metric is not known from the beginning. It has to be 
determined e.g. from a perturbative solution of the field equations which 
couple metric and matter. This calculation can be done in any reasonable 
gauge, e.g. in the harmonic gauge. Subsequently 
one can transform to our optimal gauge, thus removing all unphysical 
gauge modes.

The covariant fitting procedure presented here is, of course, not unique
 because one could modify the minimization steps which define the 
optimization.
But it is not at all trivial to formulate conditions 
which do not involve the initially unknown scale factor, can be simply 
treated perturbatively, and lead to a maximal fixing 
of the coordinates. One could e.g. exchange the
order of the steps. We did not perform a systematic
investigation of all possibilities but  chose an order which was
motivated by technical simplicity. If one would perform steps 2 and 3 at 
the beginning one 
would obtain the synchronous gauge. The E-L equations for the former step 1 
would subsequently become rather ugly. There is now no reason for 
$\langle \delta G^{(1)}_{00} \rangle $ and 
$\langle \delta G^{(1)m}_m \rangle $ to vanish, one could 
obtain a (small) backreaction already in first order.

It is worthwhile to emphasize that familiar gauge fixing prescriptions in the 
literature do not fix the gauge in the maximally possible way 
as it was achieved here. This is well known for the synchronous gauge, 
but it is also true for the Newtonian gauge, even if one restricts to 
time independent transformations. As an example consider the nonlinear
transformation $\xi_0 = 0, \; 
\xi^m = \alpha^m r^2 - 2 x^m \sum_k  \alpha^k x^k$, with $\alpha^k $
a constant vector. This leaves $g_{00}$ and $g_{m0}$ unchanged, and only 
changes
$A$ in the decomposition (3.10) to $A - 4 \sum_k \alpha^kx^k$. Therefore 
demanding e.g. $B=C_m = F =0$ is not sufficient for fixing the gauge.

We came to the conclusion that dark energy cannot be mimicked  from 
backreaction. The only quantities which agree are the signs 
$\rho_b >0$ and $p_b <0$. But the relative importance
of backreaction in the present day universe turned out to be of order 
$10^{-4}$ to at most $10^{-2}$. 
Somewhat different parameters than the ones used above, 
or more realistic density distributions inside the sources, 
would not change the numbers considerably. 
We also gave arguments why a perturbative expansion makes sense,  
and why neither the presence of large voids nor
retardation effects should seriously modify the results. Only
if one widely gives up the cosmological principle one may evade 
these conclusions, but then it becomes hard to 
derive even semiquantitative statements.

In order to mimic dark energy from backreaction the small ratio 
$\rho_b/\overline{\rho}$ should somehow increase to the observed ratio 
$\rho_{DE}/\overline{\rho} \approx 7/3$ between dark energy and matter. It
appears miraculous how this could happen. An even greater miracle would 
be needed to change the ratio $w_b = - 1/21$ to the observed 
$w_{DE} \approx -1$ for dark energy.
\\

{\bf Acknowledgement:} I thank E. Thommes for valuable discussions and 
for reading the manuscript.

\end{document}